\documentclass[twocolumn,epjc3]{svjour3}          

\smartqed  
\RequirePackage{amsmath}
\RequirePackage{amssymb}
\RequirePackage{newtxtext}
\RequirePackage{graphicx}
\RequirePackage{mathptmx}      
\RequirePackage{flushend}
\RequirePackage[numbers,sort&compress]{natbib}
\RequirePackage[colorlinks,citecolor=blue,urlcolor=blue,linkcolor=blue]{hyperref}
\RequirePackage[toc,page]{appendix}
\RequirePackage[colorlinks]{hyperref}
\RequirePackage[T1]{fontenc}

\journalname{Eur. Phys. J. C}

\begin{document}

\title{Upper limits on the amplitude of ultra-high-frequency gravitational waves from graviton to photon conversion}

\author{A.~Ejlli\thanksref{e1,addr1}
	\and
		D.~Ejlli\thanksref{addr3}
	\and 
		A. M.~Cruise\thanksref{addr2}
	\and 
		G.~Pisano\thanksref{addr1}
	\and
		H.~Grote\thanksref{addr1}
		}
		
\thankstext{e1}{Corresponding author: EjlliA@cardiff.ac.uk}

\institute{Cardiff University, School of Physics and Astronomy, The Parade, Cardiff, CF24 3AA, UK\label{addr1}
\and 
Birmingham University, School of Physics and Astronomy, Edgbaston Park Rd, Birmingham B15 2TT, UK\label{addr2}
\and
Department of Physics, Novosibirsk State University, 2 Pirogova Street, Novosibirsk, 630090, Russia \label{addr3}
}

\maketitle

\begin{abstract}
In this work, we present the first experimental upper limits on the presence of stochastic gravitational waves in a frequency band with frequencies above 1~THz. We exclude gravitational waves in the frequency bands from $\left(2.7 - 14\right)\times10^{14}$ Hz and $\left(5 - 12\right)\times10^{18}$~Hz down to a characteristic amplitude of $h_c^{\rm min}\approx6\times 10^{-26}$ and $h_c^{\rm min}\approx 5\times 10^{-28}$ at 95\% confidence level, respectively. To obtain these results, we used data from existing facilities that have been constructed and operated with the aim of detecting WISPs (weakly interacting slim particles), pointing out that these facilities are also sensitive to gravitational waves by graviton to photon conversion in the presence of a magnetic field. The principle applies to all experiments of this kind, with prospects of constraining (or detecting), for example, gravitational waves from light primordial black hole evaporation in the early universe.

\end{abstract}



\section{Introduction}\label{sec:GW'spectrum}

  With the first detections of gravitational waves (GWs) by the ground-based laser interferometers LIGO and VIRGO, a new tool for astronomy, astrophysics and cosmology has been firmly established \cite{GW_150914,GW_170817}. GWs are spacetime perturbations predicted by the theory of general relativity that propagate with the speed of light and can be predominantly characterised by their frequency $f$ and the dimensionless (characteristic) amplitude $h_c$. Based on these two quantities and the abundance of sources across the full gravitational-wave spectrum, as well as the availability of technology, it becomes clear that different parts of the gravitational-wave spectrum are more accessible than others.
 
Current ground-based detectors are sensitive in the frequency band from about 10~Hz to 10~kHz \cite{LIGO2016_InstrumentLong,VIRGO2015,GEO600_2004,KAGRA_2012} where the intersection of efforts in the development of the technology and the abundance of sources facilitated the first detections. Coalescences of compact objects such as black holes and neutron stars have been detected, and spinning neutron stars, supernovae and stochastic signals are likely future sources.
Since in principle, GWs can be emitted at any frequency, they are expected over many decades of frequency below the audio band, but also above it. At lower frequencies, the space-based laser interferometer LISA is firmly planned to cover the $0.1-10$~mHz band \cite{eLISA2014,update-egL3proposal}, targeting, for example, black hole and white dwarf binaries. At even lower frequencies in the nHz regime, the pulsar timing technique promises to facilitate detections of GWs from supermassive black holes \cite{PPTA2010,EPTA2013,NANOGRAV2013}.

Frequencies above 10 kHz have been much less in the focus of research and instrument development in the past, but given the blooming of the field, it seems appropriate to not lose sight of this domain as technology progresses. One of the main reasons to look for such high frequencies of GWs is that several mechanisms that generate very high-frequency GWs are expected to have occurred in the early universe just after the big bang. Therefore, the study of such frequency bands would give us a unique possibility to probe the very early universe. However, the difficulty in probing such frequency bands is explained by the fact that laser-interferometric detectors such as LIGO, VIRGO and LISA work in the lower frequency part of the spectrum and their working technology is not necessarily ideal for studying very-high-frequency GWs. The characteristic amplitude of a stochastic background of GWs $h_c$, for several models of GW generation, decreases as the frequency $f$ increases. Consequently, to study GWs with frequencies in the GHz regime or higher requires highly sensitive detectors in terms of the characteristic amplitude $h_c$.

One possible way to construct detectors for very-high-frequency GWs is to make use of the partial conversion of GWs into electromagnetic waves in a magnetic field. Indeed, as general relativity in conjunction with electrodynamics predicts, the interaction of GWs with electromagnetic fields, in particular, static magnetic fields, generate propagating electromagnetic radiation at the same frequency as the incident GW. In other words, gravitons mix with photons in electromagnetic fields. This effect has been studied in the literature by several authors in the context of a static laboratory magnetic field \cite{Boccaletti70,Zeldovich1973,deLogi77,Raffelt88} and in astrophysical and cosmological situations \cite{P.Macedo1983,fargion95,Cruise2012,Dolgov:2012be,Ejlli:2018hke}. The effect of graviton-photon (also denoted as GRAPH) mixing is the inverse process of photon-graviton mixing studied in Refs.~\cite{Gertsenshtein1962,Lupanov1971,Raffelt88,Bastianelli05,Bastianelli07}.

Based on the graviton-photon (or GRAPH) mixing, in this work we point out that the existing experiments that are conceived for the detection of weakly interacting slim particles (WISPs) are also GW detectors in a sense mentioned above: they provide a magnetic field region and detectors for electromagnetic radiation. In this work, we make use of existing data of three such experiments to set first upper limits on ultra-high frequency GWs.
As technology may progress further, future detectors based on the graviton to photon conversion effect may be able to reach sensitivities for GW amplitudes near the nucleosynthesis constraint at the very high-frequency regime.

This paper is organised as follows: In Sec. \ref{sec:2}, we give an overview of high-frequency GW sources and generating mechanisms as well as previously existing experimental upper limits. In Sec. \ref{sec:3} we discuss the working mechanism of current WISP detectors and the possibility to use them as GW detectors. In Sec. \ref{sec:4}, we consider the minimum GW amplitude that can be detected by current WISP detectors. In Sec. \ref{sec:5}, we discuss the prospects to detected ultra-high-frequency GWs with current and future WISP detectors and in Sec. \ref{sec:6} we conclude. In this work we use the metric with signature $\eta_{\mu\nu}=\text{diag}[1, -1, -1, -1]$ and work with the rationalised Lorentz-Heaviside natural units ($k_B=\hbar=c=\varepsilon_0=\mu_0=1$) with $e^2=4\pi \alpha$ if not otherwise specified.


\section{Overview of high frequency GW sources and detection amplitude upper limits}
\label{sec:2}

The gravitational wave emission spectrum has been fully classified from $\left(10^{-15}-10^{15}\right)$~Hz, as for example, more recently in \cite{Book_W.Ni2016}. For the frequency region of interest to this paper, the high-frequency GW bands are given as: 

\begin{itemize}
\item { High-Frequency band (HF), ($10-100$~kHz)},
\item Very High-Frequency band (VHF), (100~kHz$~-1$~THz),
\item Ultra High-Frequency band (UHF), (above 1~THz).
\end{itemize}

A viable detection scheme in the VHF and UHF bands (but in principle at all frequencies), is the graviton to photon conversion effect. Based on this effect, it seems feasible to search for GWs converted to electromagnetic waves in a magnetic field. The generated electromagnetic waves can be processed with standard electromagnetic techniques and can be detected, for example, by single-photon counting devices at a variety of wavelengths. Following the classification of high frequency source in the paper \cite{Cruise2012}, there appear to be four kinds of potential GW sources in the VHF and UHF bands: 

1) Discrete sources \cite{Kogan2004}: the authors examined the thermal gravitational radiation from stars, mutual conversion of gravitons and photons in static fields and focusing the main attention to the phenomenon of primordial black-hole evaporation, with a backgrounds at the high-frequency region.

2) Cosmological sources \cite{Grishchuk1976}: another mechanism which generates a very broadband energy density of GWs noise in the form of non-equilibrium of cosmic noise generated as a consequence of the super-adiabatic amplification at the very early universe. An upper bound on the energy density independent of the spectrum of any cosmological GWs background prediction is given from the nucleosynthesis bound of $\Omega_{\rm GW} \geq 10^{-5}$ \cite{Maggiore:1999vm}. 
 
3) Braneworld Kaluza-Klein (KK) mode radiation \cite{Clarkson2007,Servin2003}: the authors suppose the existence of the fifth dimension in higher-dimensional gravitational models of black holes derives emission of the GWs. The GWs are generated due to orbital interactions of massive objects with black holes situated on either our local, ``visible", brane or the other, ``shadow", brane which is required to stabilise the geometry. These KK modes have frequencies which may lie in the UHF frequency with large amplitudes since the gravity is supposed to be very strong in bulk with a large number of modes.
 
 4) Plasma instabilities \cite{Servin2003}: the authors have modelled the behaviour of magnetised plasma example supernovae, active galaxies and the gamma-ray burst. They have developed coupled equations linking the high-frequencies electromagnetic and gravitational wave modes. Circularly polarised electromagnetic waves travelling parallel to plasma background magnetic field would generate gravitational-wave with the same frequency of the electromagnetic wave.

 Except for the discrete sources and plasma instabilities, the GW radiation is usually emitted isotropically in all directions for several generation models, see below. Normally, a GW detector should be oriented toward the source in order to efficiently detect GWs, except for the majority of cosmological sources. Indeed, cosmological sources are expected to generate a stochastic, isotropic, stationary, and Gaussian background of GWs that in principle can be searched for with an arbitrary orientation of the detector. The upper limits on the GW amplitudes that we derive in this paper are limited to the cosmological sources since the detectors (see next section), except for one experiment which pointed towards the sun, cannot point deliberately to the emitting sources, so their measurements are most sensitive to sources creating an isotropic background of GWs. We list some possible sources of GWs:
 
\begin{itemize}

\item[1)] Stochastic background of GWs

The stochastic background of GWs is assumed to be isotropic \cite{Allen1997,Allen:1997ad,Maggiore:1999vm} and must exist at present as a result of an amplification of vacuum fluctuations of gravitational field to other mechanisms that can take place during or after inflation \cite{Grishchuk2003}. Inflationary processes and the hypothetical cosmic strings are potential candidates of the GW background with some differences in the predicted intensity and spectral features \cite{Allen1997,Allen:1997ad,Maggiore:1999vm}. 
These spectrum would have cutoff frequency approximately in the region $\nu_{c}\sim 10^{11}$ Hz. The prediction for the cutoff frequency in some cosmic string models gives the cutoff shifted to $10^{13}-10^{14}$ Hz \cite{Allen1997,Allen:1997ad,Maggiore:1999vm}. The metric perturbation at the cutoff frequency $10^{11}$~Hz corresponds to an estimated strain amplitude of $h_c\sim 10^{-32}$.\\

\item[2)] GWs from primordial black holes

In Ref. \cite{Nakaruma1997}, the authors proposed the existence of Primordial Black Hole (PBH) binaries and estimated the radiated GW spectrum from the coalescence of such binaries. In addition, the mechanism of evaporation of small mass black holes gives rise to the production of high and ultra-high frequency GWs. An estimation of the efficiency of this emission channel which might compensate the deficit of high-frequency gravitons in the relic GW background has been thoroughly studied in \cite{Kogan2004}. A detailed calculation of the energy density of relic GWs emitted by PBHs has been performed in \cite{Ejlli2011}. The author's analysed and calculated the energy density of GWs from PBH scattering in the classical and relativistic regimes, PBH binary systems, and PBH evaporation due to the Hawking radiation. \\

\item[3)] GWs from thermal activity of the sun

A third class that does generate a stochastic, but not an isotropic background of GWs, which is relevant to this work, is the GW emission from the sun \cite{CAST2017}. The high temperature of the sun in the proton-electron plasma produces isotropic gravitational radiation noise due to thermal motion \cite{Braginskii1963,Weinberg1972,Galtsov1974}. The emission comes to the detector from the direction of the sun, and the observations have the potential to set limits on this process. The frequency of the collisions of $\nu_{c}\sim  10^{15}$~Hz determines the gravitational wave frequency, and the highest frequency corresponds to the thermal limit at $ \omega_{m}=kT/\hbar \sim 10^{18}$ Hz. Using the plasma parameters in the centre of the Sun the estimation of the ``thermal gravitational noise of the Sun'' reaching the earth provides a stochastic flux at ``optical frequencies'' of the order $h_c\sim 10^{-41}$ \cite{Weinberg1972}.\\ 
\end{itemize}


So far, dedicated experiments to detect GWs in the VHF region are based on two designs: polarisation measurement on a cavity/waveguide detector and cross-correlation measurement of two laser interferometers. The cavity/waveguide prototype measured polarisation changes of the electromagnetic waves, which in principle can rotate under an incoming GW, providing an upper limit on the existence of GWs background to a dimensionless amplitude of $h_c\leq 1.4\times10^{-10}$ at 100~MHz \cite{Cruise2006}. The two laser interferometer detectors with $0.75$~m long arms have used a so-called synchronous recycling interferometer and provided an upper limit on the existence of the GW background to a dimensionless amplitude of $h_c\leq1.4\times10^{-12}$ at 100 MHz \cite{Akutsu2008,Nishizawa2008}.

The most recent upper limit on stochastic VHF GWs has been set from by a graviton-magnon detector which measures resonance fluorescence of magnons \cite{Jiro2019}. They used experimental results of the axion dark matter using magnetised ferromagnetic samples to derive the upper limits on stochastic GWs with characteristic amplitude: $h_c \leq 9.1\times10^{-17}$ at 14~GHz and $h_c \leq 1.1\times10^{-15}$ at 8.2~GHz.

Another facility, the Fermilab Holometer, has performed measurement at slightly lower frequencies. The Fermilab Holometer \cite{Holo2017} consists of separate, yet identical Michelson interferometers, with $39$~m long arms. The upper limits found within 3$\sigma$, on the amplitude of GWs are, in the range $h_c < 25\times10^{-19}$ at 1~MHz down to a $h_c < 2.4\times10^{-19}$ at 13~MHz.

\section{WISP search experiments and their relevance to UHF GWs}
\label{sec:3}

\begin{figure}[htbp]
\begin{center}
\includegraphics[width=8.6cm]{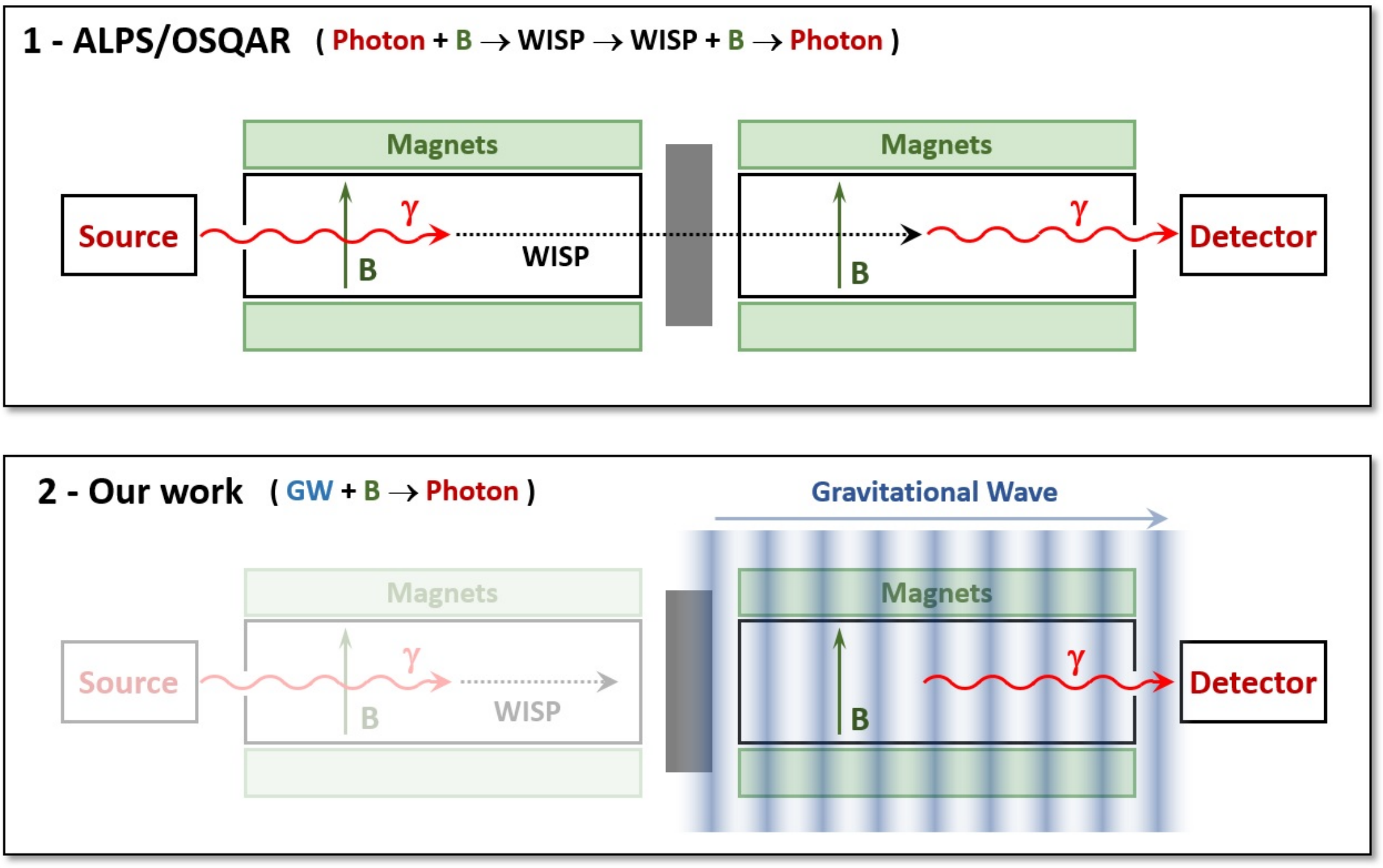}
\caption{Simplified schematic of the experimental setup aiming at the detection of WISPs. In the upper panel left-hand side, the electromagnetic waves interacting with the magnetic field produce the hypothetical WISPs, and at the right-hand side electromagnetic waves are produced by the decay of WISPs in the constant magnetic field. Our work is illustrated by the lower panel ignoring the transparent left-hand side. On the right hand side, the photons detected could be due to the passage of GWs propagating in the constant magnetic field.}
\label{fig:setup_alps}
\end{center}
\end{figure}

\begin{table*}[htbp]
\centering
\begin{tabular}{|c|c|c|c|c|c|c|}
\hline
\hline
        	 		& $\epsilon_{\gamma}\left(\omega\right)$              & 	$N_{\rm exp}$~(mHz) 		&	$A$~(m$^2$)	& $B$~(T) & $L$~(m) & $\Delta f$~(Hz) \\ \hline
{\rm ALPS}~I     		& Fig~\ref{fig:QE} 	&	  			0.61		&          $0.5\times10^{-3}$ & 	5     			& 9 & $ 9\times10^{14}$\\ \hline
{\rm OSQAR}~I  	& Fig~\ref{fig:QE} 		&   			1.76		&    $0.5\times10^{-3}$ & 	9     			& 14.3 &$5\times10^{14} $\\ \hline
{\rm OSQAR}~II 	& Fig~\ref{fig:QE} 		&  			1.14		&         $0.5\times10^{-3}$   		& 	9     			& 14.3 &$1\times10^{15}$ \\ \hline
{\rm CAST}     		& Fig~\ref{fig:QE} 		&   			0.15		& 	$2.9\times10^{-3}$      	& 	9     			& 9.26  & $1\times10^{18}$\\ \hline
\end{tabular}
\caption{Relevant characteristics of the experimental setups, as operated for the detection of WISPs, and used for GW upper limits in this work. The reported quantities are used to estimate the minimum detectable GW amplitude through the graviton to photon conversion in a constant and transverse magnetic field.}
\label{tab:1}
\end{table*}

The experiments ALPS \cite{ALPS2010}, OSQAR \cite{OSQAR2015} and CAST \cite{CAST2017} have not been designed to detect GWs in the first place. However, in this work their results are used to compute new upper limits on GW amplitudes and related parameters. 
The experiments performed by ALPS and OSQAR are usually called ``light shining through the wall experiment'' where the hypothetical WISPs, that are generated within the experiment, mediate the ``shining through the wall'' process, and decaying successively into photons. In contrast, the CAST experiment searches directly for WISPs emitted from the core of the sun. Though all these experiments are not designed to detect gravitational waves, they provide a high sensitivity measurement of single photons generated in their constant magnetic field which is the crucial ingredient for the detection of graviton to photon conversion.
 
 The main characteristics of the ALPS, OSQAR and CAST experiments are:
\begin{itemize}
\item[1)] ALPS experiment at DESY

The ALPS (I) experiment has performed the last data taking run in 2010, and the specific characteristics of the experiment are found in Ref. \cite{Ehret:2009sq}. A general schematic of the principle is shown in the upper panel of the Fig. \ref{fig:setup_alps}. The production of WISPs and their re-conversion into electromagnetic radiation is located in one single HERA superconducting magnet where an opaque wall in the middle separates the two processes. The HERA dipole provides a magnetic field of $5$~T in a length of $8.8$~m. The electromagnetic radiation, generated by the decay of the WISPs in the magnetic field, passed a lens of 25.4~mm diameter and focal length $40~$mm. The lens focuses the light onto a $\approx 30~\mu$m diameter beam spot on a CCD camera.\\

\item[2)] OSQAR experiment at CERN

The OSQAR experiment has performed the last data taking in 2015, and the specific characteristics of the experimental setup are found in \cite{OSQAR2014,OSQAR2015}. The OSQAR collaboration has used two LHC superconducting dipole magnets separated by an optical barrier, (for a conceptual scheme see the upper panel of the Fig.~\ref{fig:setup_alps}). The LHC dipole magnets provide a constant magnetic field of 9~T, along a length of 14.3~m. To focus the generated photons of the beam onto the CCD, an optical lens of 25.4~mm diameter and a focal length of 100~mm was used, installed in front of the detector similar to the Fig.~\ref{fig:setup_alps}.  Data acquisition has been performed in two runs with two different CCD's having different quantum efficiencies.\\

\item[3)] CAST experiment at CERN 

The CERN Axion Solar Telescope (CAST) experiment has the aim to detect or set upper limits on the flux of the hypothetical low-mass WISPs produced by the Sun. A refurbished CERN superconducting dipole magnet of 9~T and 9~m length was used. The solar Axions with expected energies in the keV range can convert into X-rays in the constant magnetic field, and an X-ray detector has been used to performed runs in the time period 2013 - 2015 \cite{Javier2015}. To increase the cross section, both the two parallel pipes which pass through the magnet have been used which provide an area of $2\times14.5$~cm$^2$ focused into a Micromegas detector.  The CAST detector mounted on the pointing system had a telescope with a focal length of 1.5~m installed for the $\left(0.5-10\right)$~keV energy range. 
\end{itemize}

\section{Minimal-detection of GW amplitude}\label{sec:4}

\begin{figure}[htbp]
\begin{center}
\includegraphics[width=8.6cm]{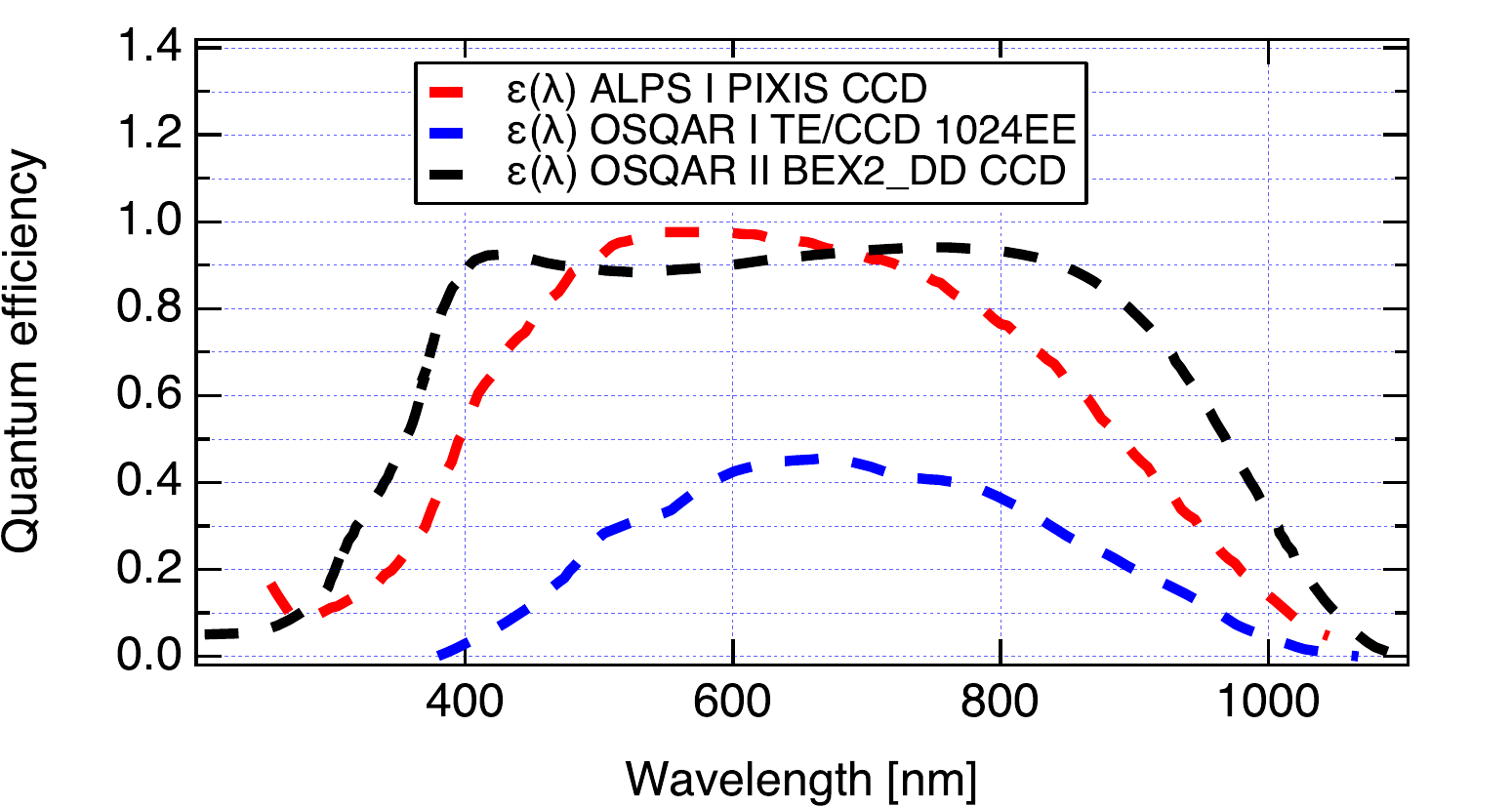}
\includegraphics[width=8.6cm]{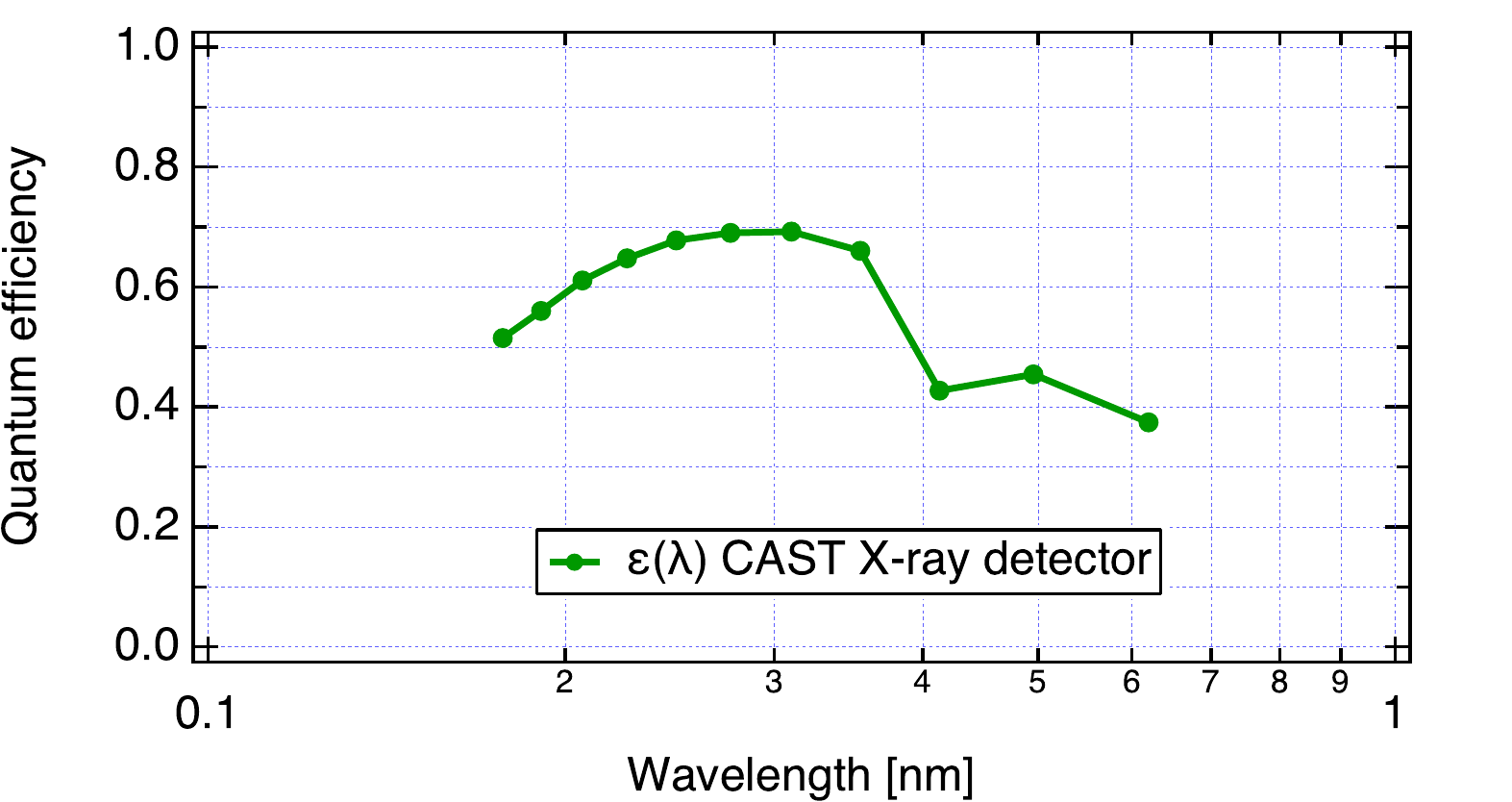}
\caption{Quantum efficiency as a function of the wavelength. Left-hand side panel: the quantum efficiency of the detectors using the method ``light shining through a wall''; in the right-hand panel: the quantum efficiency of the Micromegas X-ray detector used in the CAST experiment. The detector bandwidth and their normalised quantum efficiency function are used to compute the upper limits un GWs detectors.}
\label{fig:QE}
\end{center}
\end{figure}

In this section, we show how we compute upper limits on the GW dimensionless amplitude $h_c$ based on the characteristics of the experiments described above that are sensitive to an isotropic background of GWs from cosmological sources and to the thermal activity in the sun. We ignore the generation of WISPs (Fig. \ref{fig:setup_alps}: lower panel) and focus on the second half of the magnetic field for the case of ALPS and OSQAR experiments, and, for the CAST experiment we consider the full magnet region. These experiments measure a number of photons per unit time with their CCD detectors, namely $N_{\rm exp}$ in an energy band $\Delta\omega$ with efficiency $\epsilon_{\gamma}$ and in a cross-section $A$. In what follows we assume that in the CCD, the background dark current fluctuation is a stochastic process with uniform probability distribution and stationary in $\omega$. In this case, the energy flux of photons generated in an energy bandwidth $[\omega_i, \omega_f]$ is given by:

\begin{equation}\label{Flux_CCD}
\Phi_\gamma^\text{CCD}\left(z, \omega_f; t\right)= \int_{\omega_i}^{\omega_f} {1\over A\left(z\right)}\frac{N\left(\omega, t\right) \,\omega}{\epsilon_{\gamma}\left(\omega\right)}\,d\omega 
\end{equation}
where $N\left(\omega,t\right)$ is the number of photons per unit of time and energy. Now, we have to compare the measured energy flux of photons with the intrinsic energy flux of photons generated in the graviton to photon conversion in presence of an isotropic background of GWs. The analytical treatment for an isotropic background of GWs converted into electromagnetic waves, is described in detail in Appendix~\ref{sec:2}, \ref{sec:3}. In Appendix \ref{sec:3}, different useful quantities are calculated, for stochastic GWs propagating in a transverse and constant magnetic field. Since all the experiments operated under vacuum condition and the propagation distance $z$ is small with respect to the oscillation length of the particles, we can safely take $\Delta_{x, y}\,z \ll 1$. The variable $\Delta_{x, y}$ defined in the Eq.~\ref{delta-def} is a function of the magnetic field, the magnitude of the photon and graviton wave-vectors and Newtonian constant. Then the energy flux of photons generated in the magnetic field of length $z$ given by expression \eqref{en-log-int}, in the same energy bandwidth $[\omega_i, \omega_f]$ becomes:

\begin{eqnarray}
 \nonumber \Phi_\gamma^\text{graph}\left(z, \omega_f; t\right) &=&\left(M_{g\gamma}^x\right)^2\, \int_{\omega_i}^{\omega_f }  \left[ \frac{ \sin^2\left(\Delta_x z\right)} {\Delta_x^2} + \frac{  \sin^2\left(\Delta_y z\right)} {\Delta_y^2} \right]\times\\
 \nonumber&\times& \frac{h^2_c\left(0, \omega\right)\, \omega}{2~\kappa^2} d\omega \\
 &\simeq& \int_{\omega_i}^{\omega_f}\frac{B^2\, z^2\, h_c^2\left(0, \omega\right)\, \omega}{4}  
\label{Flux}
\end{eqnarray}

Comparing the energy fluxes in expressions \eqref{Flux_CCD} and \eqref{Flux}, and requiring that for detection, the energy flux in \eqref{Flux} must be bigger or equal to the energy flux in \eqref{Flux_CCD}, we get

\begin{equation}\label{final-hc}
h_c^2\left(0, \omega\right) \geq  \frac{4\, N\left(\omega, t\right)}{B^2 \,L^2\, A\left(L\right) \,\epsilon_\gamma\left(\omega\right)},
\end{equation}
where we took $z=L$ with $L$ being the spatial extension of the external magnetic field. All the three experiments listed in the section above their upper limits are compatible to the background fluctuation of the detector which allows us to express the relation: $N \left(\omega, t\right)=N_{\rm exp}/\Delta \omega$, where $\Delta\omega= \omega_f-\omega_i$. Finally, by putting the units in explicitly, we get the following expression for the minimum detectable GW amplitude:

 \begin{eqnarray}
\nonumber &&h^{\rm min}_c\left(0, \omega\right) \simeq \sqrt{4\, N_{\rm exp}\over A\, B^2\, L^2 \,\epsilon_{\gamma}\left(\omega\right)\, \Delta\omega } \simeq 1.6\times10^{-16}\times \\
&\times&\left({1~{\rm T}\over B}\right)\left({ 1~{\rm m}\over L}\right)\sqrt{\left({N_{\rm exp}\over 1~\rm{Hz}}\right) \left({1~{\rm m}^2 \over A}\right) \left({1~{\rm Hz}\over \Delta f}\right) \left({1\over\epsilon_{\gamma}}\right)}
\label{min_amplitude}
\end{eqnarray}

where $\omega=2\pi f$ with $ f$ being the frequency. In order to compute the minimum detectable GW amplitude, $h_c^\text{min}$, we have extracted the following quantities from the ALPS, OSQAR and CAST experiments: 
 
 \begin{itemize} 
 \item $N_{\rm exp}$ the total detected number of photons per second in the bandwidth $\Delta \omega$, 
 \item $A$ cross-section of the detector, 
 \item $B$ magnetic field amplitude, 
 \item $L$ distance extension of the magnetic field, 
 \item $\epsilon_{\gamma}(\omega)$ quantum efficiency of the detector, 
 \item $\Delta f$ operation bandwidth of the CCD.
 \end{itemize} 
 
These quantities permit to compute the equivalent minimum amplitude $h^{\rm min}_c$ of a stochastic GW background which would generate photons through graviton to photon conversion, equivalent number of background photons that the CCD has read. The data accounted for the photon detection in the constant magnetic field for the three experiments ALPS, OSQAR and CAST, exclude the detection of physical signals with fluxes bigger or comparable to the background count of the CCD detectors at 95\% confidence level which allows putting upper limits on the minimal detectable GW amplitude $h_c^\text{min}$ at the same confidence level.

The extracted quantities used to compute the upper limits on $h^{\rm min}_c$ are summarised in Table \ref{tab:1}. These experiments attempting to detect WISPs have used subsequently improved CCDs during different run phases which is taken into account in the analysis. The quantum efficiencies in Table~\ref{tab:1} are represented graphically in Fig.~\ref{fig:QE} as a function of the wavelength. We have taken into account that $N_{\rm exp}$ is normalised to the quantum efficiency the working frequency of the WISPs experiments, and the range of the expected photons is imposed by the sensitive frequency range of the CCD.

The cross-section reported in Table \ref{tab:1} has been computed, for the ALPS and OSQAR experiments, considering the area enclosed by the diameter of the lens \cite{ALPS2010,OSQAR2015}. Instead, the CAST experiment uses the whole cross-section of the two beam pipes.

\begin{figure}[htbp]
\begin{center}
\includegraphics[width=8.6cm]{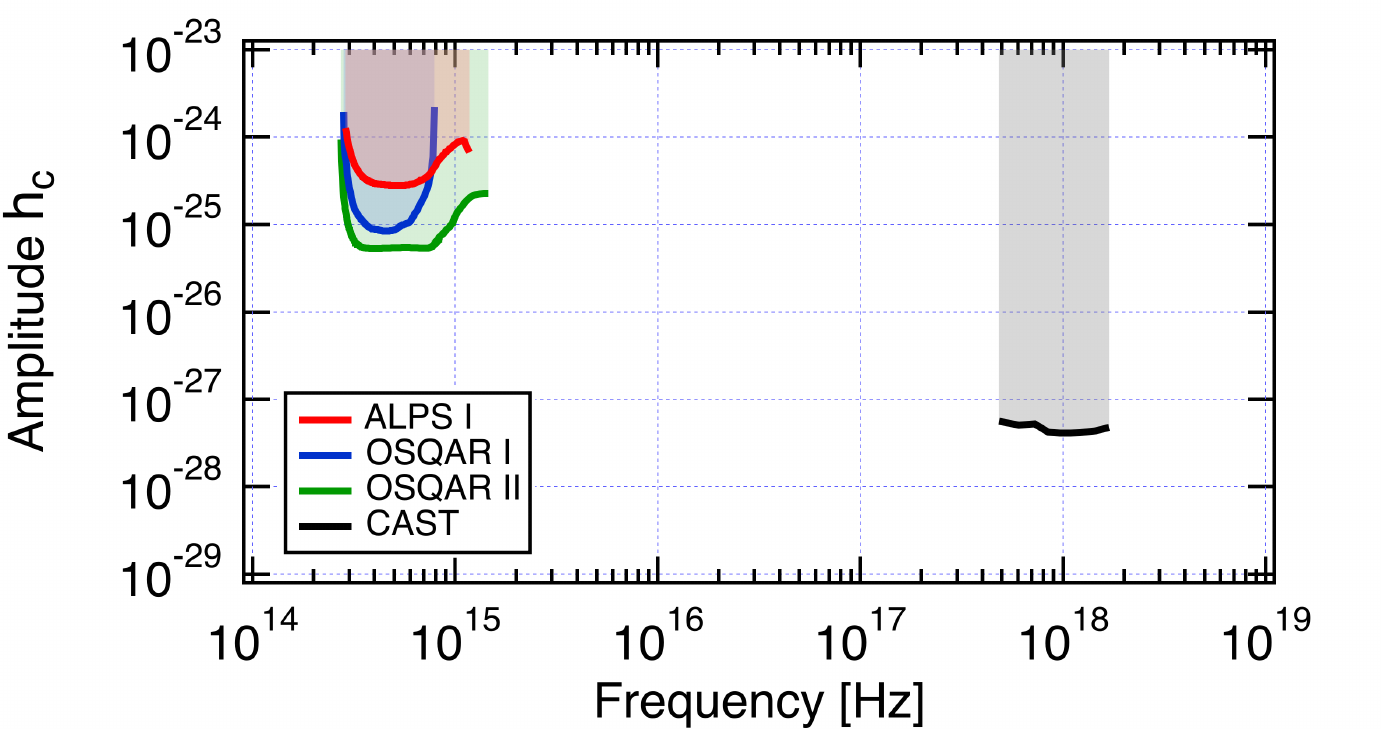}
\caption{Plots of the minimum detectable GW amplitude $h_c^\text{min}$ as a function of the frequency $f$, deduced from the measured data of the denoted experiments.}
\label{fig:ampl}
\end{center}
\end{figure}

Using the data of the Table~\ref{tab:1} and expression \eqref{min_amplitude} it is possible to produce an upper limit plot of the GW amplitude, see Fig.~\ref{fig:ampl}, due to the conversion of GWs into photons. The region above each curve is the excluded region. To our knowledge, these are the first experimental upper limits in these frequency regions.


\section{Prospects on detecting Ultra-High Frequency GWs from primordial black holes}
\label{sec:5}

  \begin{figure}[htbp]
\begin{center}
\includegraphics[width=8.6cm]{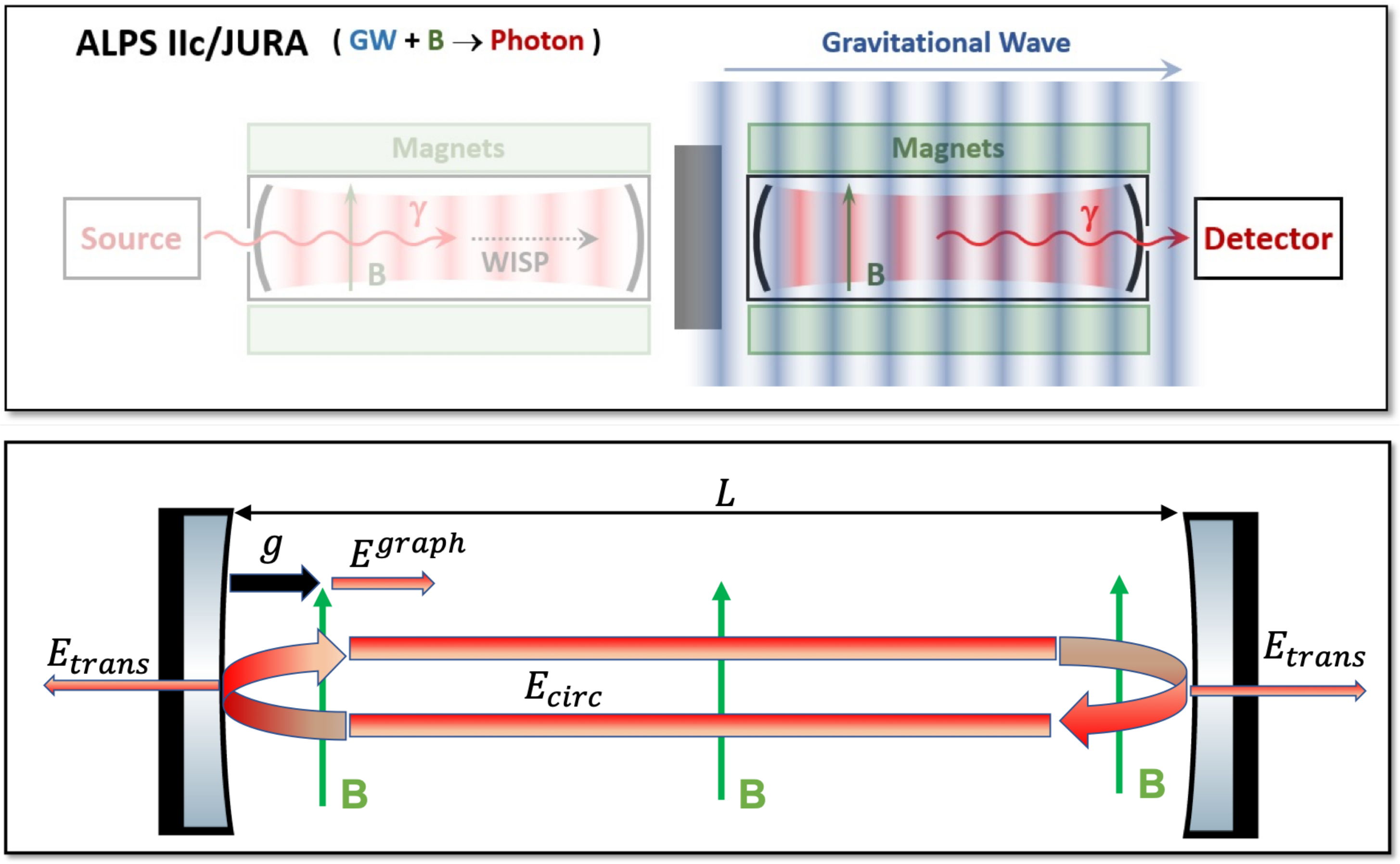}
\caption{In the upper panel conceptual scheme of the experimental setup ALPS IIc and a possible follow-up named JURA where we note the addition of the FP cavity in the right-hand side. Our prediction for the sensitivity of the minimal amplitude of $h^{\rm min}_c$ used the right-hand side process, where the photons generated via graviton to photon conversion are resonantly enhanced in the FP cavity. In the lower panel, the FP resonator concept is described where $E^{\rm graph}$ is the electric field generated from the graviton to photon conversion in the cavity, $E_{\rm circ}$ is circulating electric field accumulated inside the resonator after transmission losses on both mirrors, $E_{\rm trans}$ is transmitted electric field through the mirrors and $L$ the length of the cavity.}
\label{fig:setup_alps_follow-up}
\end{center}
\end{figure}

 Graviton to photon conversion maybe a useful path towards the detection of UHF GWs. The actual technology has made further progress in the detection of single photons and new facilities are intended for WISP search, using higher values of $B$ and $L$ in order to achieve higher sensitivities. One facility which plans to do so is the ALPS IIc proposal which consists of two $120$~m long strings of 12 HERA magnets each, with a magnetic field of 5.3~T. The scheme of generation and conversion of the WISPs is still the same, expected an optical resonator is added to the reconversion region. A follow-up of the CAST telescope is the proposed International Axion Observatory (IAXO). Tab.~\ref{tab:2} represent the detector parameters of ALPS IIc \cite{ALPSII}, a possible follow-up named JURA \cite{JURA}, and of the IAXO proposal \cite{IAXO}.
  
\begin{table*}[htbp]
\centering
\label{tab:follow-up}
\begin{tabular}{|c|c|c|c|c|c|c|c|}
\hline
\hline
             &  $\epsilon_{\gamma}$&   $N_{\rm dark}$~(Hz)  &      $A$~(m$^2$)        & $B$~(T)         &  $L$~(m)  & $\cal{F}$ \\ \hline
ALPS IIc     &     0.75            &       $\approx 10^{-6}$       &      $\approx2\times10^{-3}$ &   5.3           &  120      & $40\,000$ \\ \hline
JURA         &      1              &       $\approx 10^{-6}$        &      $\approx8\times10^{-3}$  &   13            &  960      & $100\,000$ \\ \hline
IAXO         &      1              &       $\approx 10^{-4}$        &      $\approx21$  			&   2.5            &  25      & -  \\ \hline

\end{tabular}
\caption{Parameters of ALPs IIc, JURA and IAXO proposals used to estimate the predicted minimum detectable GW amplitude through the graviton to photon conversion in their constant and transverse magnetic field: $\epsilon_{\gamma}$ is the efficiency photodetector at 1064~nm, $N_{\rm dark}$ correspond to the number of photons per unit of time limited by the dark count sensitivity, $A$ is the cross-section, $B$~(T) is the magnetic field magnitude, $L$ is the magnetic field length and $\cal{F}$ is the finesse of the cavity.}
\label{tab:2}
\end{table*}

Since the working frequency of the detectors is different we compute the sensitivity to detected GWs, with the graviton-photon mixing process, in two frequency regions infrared and X-rays:

  
\subsection{Infrared}
  
   One of the most important changes that ALPS IIc, with respect to the ALPS I and OSQAR, is the use of a Fabry-Perot (FP) cavity to enhance the decay processes of WISPs into photons, see Fig.~\ref{fig:setup_alps_follow-up}. The FP cavity will allow just a range of electromagnetic waves to be built up resonantly, within the cavity bandwidth: $\Delta \omega_c = {\Delta \omega_{\rm FSR} / {\cal F}}$ where ${\cal F} = \pi / \left(1-R\right)$ is the cavity finesse, $\Delta \omega_{\rm FSR}= \pi/L$ is the cavity free spectral range, and $R$ is the reflectance of the mirrors. The FP cavity enhances the decay rate of WISPs to photons \cite{Sikivie2007}. This is an essential aspect because it will also account for the transition of gravitons into photons \cite{Rudenko2015}. Stochastic broadband GWs converted into electromagnetic radiation would excite several resonances of the cavity at frequencies $\omega_{\rm c}\pm n\Delta \omega_{\rm FSR}$, where $\Delta \omega_c$ is the cavity frequency bandwidth, and $n$ is an integer number with its range depending on the coating of the mirrors. To calculate the response of the FP resonator, we use of the circulating field approach \cite{Siegman1986,Nur_Ismail2016}, as displayed in the lower panel of Fig.~\ref{fig:setup_alps_follow-up}. We assume a steady state approximation to derive the circulating electric field $\vec{E}_{\rm circ}$ inside the cavity and the mirrors have the same reflectance $R$ and transmittance $T$. Defining the phase shift after one round trip $2\phi\left(\omega\right)=2\omega L$, the accumulated electric field $\vec{E}_{\rm circ}$ after a large number of reflections (which can be assumed infinite in the calculations below) of the electric field $\vec{E}^{\rm\,graph}$ generated in the GRAPH mixing is: 
   
  \begin{eqnarray}
\nonumber \vec{E}^{\rm circ}_{x,y}\left(z,t\right) &=&  \vec{E}^{\rm\,graph}_{x,y}\left(z, t\right)\\
\nonumber&\times& \left( 1+R e^{-i 2\phi\left(\omega,L\right)}+\left(R e^{-i2\phi\left(\omega,L\right)}\right)^2 +\cdots \right)\\
                   \nonumber     &=&  \vec{E}^{\rm\,graph}_{x,y}\left(z, t\right)\sum_{n=0}^{\infty} \left( R e^{- i2\phi\left(\omega,L\right)}\right)^n\\
                        &=& \vec{E}^{\rm\,graph}_{x,y}\left(z, t\right) {1\over 1-R\, e^{-i 2\phi\left(\omega,L\right)}}.
 \end{eqnarray}
The circulating flux, in a time $t>\tau$ where $\tau={\cal F}L/\pi$ is the charge time of the cavity, at a distance $z=L$ is $\Phi^{\rm\,circ}_\gamma\left(L,t\right) \equiv \langle |E_x^\text{circ}\left(z, t\right)|^2\rangle + \langle |E_y^\text{circ}\left(z, t\right)|^2 \rangle$. By expanding $\vec{E}^{\rm\,graph}_{x,y}\left(z, t\right)$ as a Fourier integral and doing the same steps to derive the flux as shown in Appendix \eqref{Appendix:A} from Eq.~\ref{I-stokes} to Eq.~\ref{I-stokes-1}, then the circulating flux simplifies to the following expression:

\begin{eqnarray}
\nonumber\Phi_\gamma^\text{circ}\left(L, t\right) &\simeq& \int_{0}^{+\infty}{1\over\left(1-R\right)^2 + 4R \,\sin^2 \left[\phi\left(\omega,L\right)\right]}\\
&\times&\frac{B^2\, L^2\, h_c^2\left(0, \omega\right)\, \omega}{4} d\omega.
\label{Flux-1}
\end{eqnarray} 
where we have consider that the propagation distance $z=L$ is small with respect to the oscillation length of the particles, and we can safely take $\Delta_{x, y}z \ll 1$.  Taking the differential of the circulating flux in \eqref{Flux-1} for a given energy interval $[\omega_i, \omega]$, we find the the following relation

\begin{equation}
d\Phi^{\rm\,circ}_\gamma\left(L,\omega; t\right)= {d\Phi^{\rm\,graph}_\gamma\left(L,\omega; t\right)\over \left(1-R\right)^2 + 4R \,\sin^2 \left[\phi\left(\omega,L\right)\right]}.
 \end{equation}
where $d\Phi^{\rm\,graph}_\gamma\left(L,\omega;t\right)$ is the differential of Eq.~\ref{I-stokes-1} in a given energy interval $[\omega_i, \omega]$, which correspond to the flux of photons generated through the graviton to photon conversion without the cavity. Now we can derive the gain factor, namely $\Gamma_{\rm circ}\left(\phi\right)$, as the differential of the circulating energy flux in the resonator relative to the differential of the energy flux generated in the graviton to photon conversion without the cavity

 \begin{equation}
 \Gamma_{\rm circ }\left(\phi\right)={d\Phi^{\rm\,circ}_\gamma \left(L,\omega;t\right)\over d\Phi^{\rm\,graph}_\gamma\left(L,\omega;t\right)}= {1\over \left(1-R\right)^2 + 4R \,\sin^2 \left[\phi\left(\omega,L\right)\right]}.
 \end{equation}

For a given length $L$ and for frequencies matching the cavity resonance, or $\phi\left(\omega,L\right)=n\pi$ where $n$ is a positive integer, the internal gain enhancement factor is maximum: $\Gamma_{\rm circ}\left(n\pi\right)=\left( {\cal F}/\pi\right)^2$. In the same way, we derive the transmitted gain on both sides of the cavity, which expression is given by:

 \begin{equation}
 \Gamma_{\rm trans }\left(\phi\right) = {d\Phi^{\rm\,trans}_\gamma \left(L,\omega;t\right)\over d\Phi^{\rm\,graph}_\gamma \left(L,\omega;t\right)}= {1-R\over \left(1-R\right)^2 + 4R\, \sin^2 \left[\phi\left(\omega,L\right)\right]}.
 \end{equation}
 
Unlike before, the transmitted flux from the cavity will exhibit transmitted light peaks which gain factor, for frequencies matching the cavity resonance condition, reduces to: $\Gamma_{\rm trans}\left(n\pi\right)=\left( {\cal F}/\pi\right)$. Now we can write explicitly the equation of the flux produced from graviton to photon conversion transmitted from the cavity in a energy bandwidth $[\omega_i, \omega_f]$: 

\begin{eqnarray}
\label{Flux-cavity}
\Phi_\gamma^\text{trans}\left(L,\omega_f;t\right) &=& \int_{\omega_i}^{\omega_f}  \Gamma_{\rm trans }\left(\phi\right) \,d\Phi^{\rm\,graph}_\gamma \left(L,\omega;t\right)\\
 \nonumber &=&\int_{\omega_i}^{\omega_f} \frac{B^2_x L^2 h_c\left(0,\omega\right)^2}{4}\, \Gamma_{\rm trans}\left(\phi\right)\,\omega \,d\omega.
\end{eqnarray}

     A photo-detector placed at the transmission line of the cavity will measures an energy flux, within its bandwidth $\Delta \omega$ defined in Eq.~\ref{Flux_CCD}. According to the previous discussion, a cavity of length $L$ will transmit light peaks for frequencies $\omega^*=n\pi/L$, and such frequency should be in the interval bandwidth $\Delta \omega$. Reminding that the flux in expressions \eqref{Flux-cavity} is calculated for a stochastic process and taking into account the bandwidth of the photodetectors of ALPS IIc and JURA such condition is satisfied. Now, considering the energy flux of a photodetector limited by the dark count rate $N_{\rm dark}$, where $N\left(\omega, t\right)=N_{\rm dark}/\Delta\omega$, and comparing with the energy fluxes in expressions \eqref{Flux-cavity}, and solving for $h_c^{\rm}\left(0,\omega\right)$ in SI units, at the maximum transmission $\omega^*=n\pi/L$, becomes:
     
\begin{eqnarray}
 h^{\rm min}_c\left(0, \omega^*\right) &\simeq& 2.8\times10^{-16} \left({1~{\rm T}\over B}\right) \left({ 1~{\rm m}\over L}\right)\times \\
\nonumber&\times& \sqrt{\left({1\over{\cal F}}\right)\left({N_{\rm dark}\over 1~\rm{Hz}}\right)\left({1~{\rm m}^2 \over A}\right)\left({1~{\rm Hz}  \over \Delta f}\right) \left({1\over\epsilon_{\gamma}}\right)}.
\label{eq:GW'sAmp2}
\end{eqnarray}  
From the above equation, we can observe that to compute the sensitivity in amplitude $h_c\left(0,\omega\right)$, with respect to the case without cavity Eq.~\ref{min_amplitude}, in addition, we need to know the finesse factor ${\cal F}$. The minimum detectable GW amplitude, considering the photodetector background rate in one second, limited by the dark counting rate, a frequency bandwidth $\Delta f\approx4\times10^{14}$~Hz \cite{Jan2014}, and the relevant characteristic of Tab.~\ref{tab:follow-up}, the sensitivity for the minimal amplitude is: $h_c^{\rm ALPS~IIc}\approx 2.8 \times 10^{-30}$ and $h_c^{\rm JURA}\approx 2 \times 10^{-32}$, which is two orders of magnitude better than in the case without cavity.

\subsection{X-rays}

 The core element of IAXO will be a superconducting toroidal magnet, and the detector will use a large magnetic field distributed over a large volume to convert solar axions into detectable X-ray photons. The central component of IAXO is a superconducting toroidal magnet of 25~m length and 5.2~m diameter. Each toroid is assembled from eight coils, generating 2.5 T in eight bores of 600~mm diameter. The X-ray detector would be an enhanced Micromegas design to match the softer $1-10$~keV spectrum. The X-rays are then focused at a focal plane in each of the optics read by pixelised planes with a dark current background level of $N_{\rm dark}=10^{–4}$ \cite{IAXO}. For the process of graviton to photon conversion, the sensitivity on the minimal amplitude of $h^{\rm IAXO}_c \approx 1\times10^{-29}$. 


\subsection{Ultra-High Frequency GWs from Primordial Black Holes}
\begin{figure}[htbp]
\begin{center}
\includegraphics[width=8.6cm]{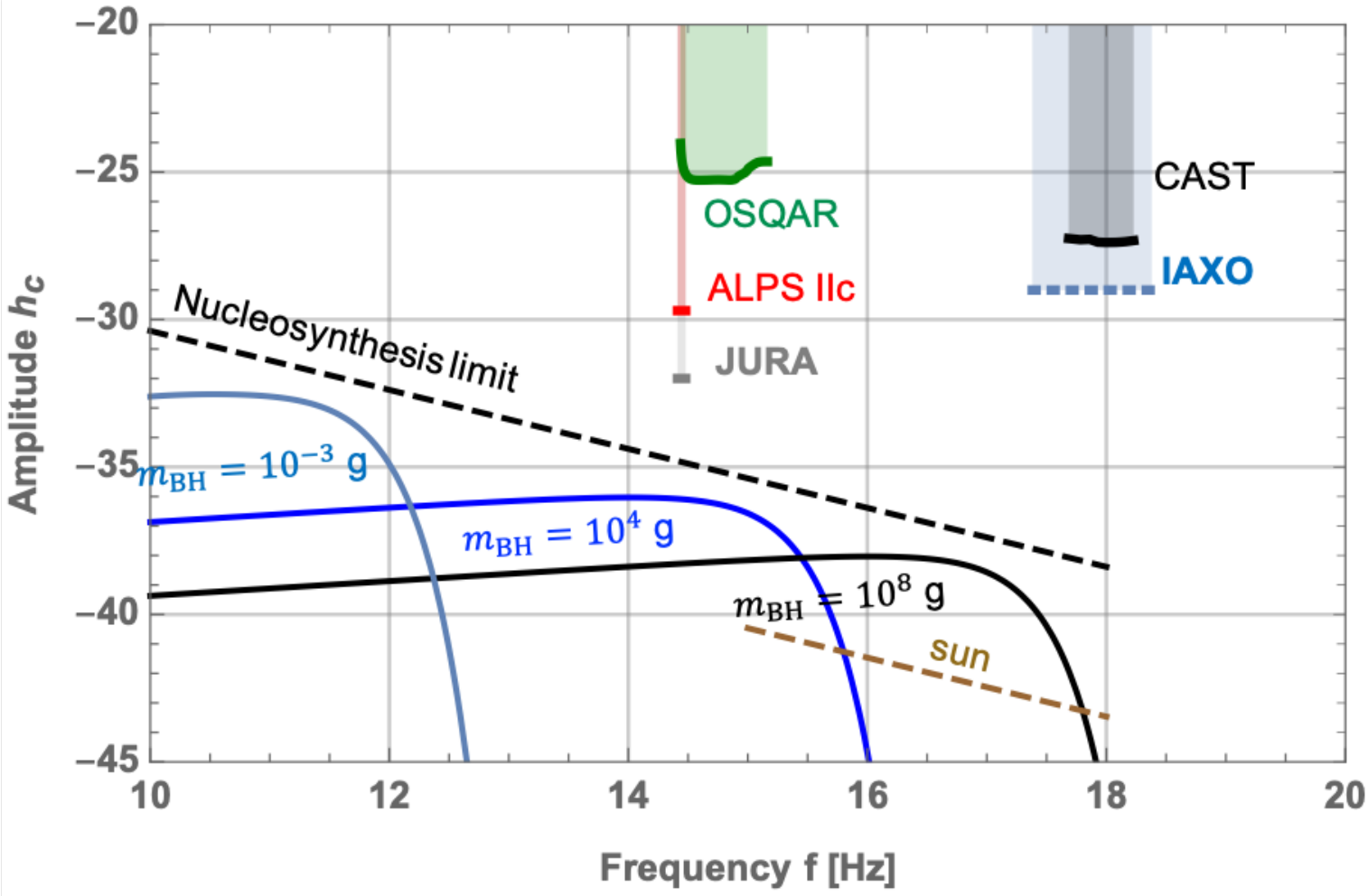}
\caption{Graphical representation of the amplitude $h^{\rm min}_c$ as a function of frequency for the: PBH evaporation of masses $m_{\rm BH}$ ($10^{-3}$, $10^4$~g, $10^8$~g), the upper limits of graviton-photon conversion data in Fig.~\ref{fig:ampl}, the estimated amplitude sensitivity for the ALPS IIc and JURA (red and grey lines) at the infrared region using their detection scheme. The dotted blue line is the estimated sensitivity for the solar telescope JAXO successors of CAST experiment. The two dashed lines represent the nucleosynthesis amplitude upper limit and the predicted amplitude from the thermal GW emitted from the sun. Here for simplicity we have assumed a value of the PBH density parameter at their production times equal to $\Omega_p\simeq 10^{-7}$. Both amplitude, $h_c$, and frequency axes are in Log$_{10}$ units.}
\label{fig:BH-Evap}
\end{center}
\end{figure}

In order to describe the potential of ALPS IIc and JURA on probing the GW background at very high frequencies an explicit example of a GW source can be considered. One of the most promising sources of VHF and UHF GWs in the frequency range of interest regarding ALPS IIc is the evaporation of very light PBHs that would have been formed just after the big bang. As shown in detail in Ref. \cite{Ejlli2011}, these black holes would emit GWs by different mechanism as scattering, binary black hole, and evaporation by hawking radiation which in principle could contribute to the spectrum of cosmic electromagnetic X-ray background due to graviton-photon mixing in cosmic magnetic fields \cite{Dolgov:2012be}. It is especially the evaporation of GWs due to Hawking radiation which generates a substantial amount of GWs in the frequency regime compatible with the ALPS IIc and JURA working frequency. The spectral density parameter of GWs at present is given in Ref. \cite{Ejlli2011} and it reads

\begin{eqnarray}
\nonumber h_0^2\Omega_\text{gw}\left(f; t_0\right)&=&1.36\times 10^{-57} \left(\frac{N_\text{eff}}{100}\right)^2 \left(\frac{1\,\text{g}}{m_\text{BH}}\right)^2 \left(\frac{f}{\text{1 Hz}}\right)^4 \\
 &\times& \int_0^{z_{\rm max}} \frac{\sqrt{1+z}}{e^{\left(\frac{2\pi f \left(1+z\right)}{T_0}\right)} -1} \,d z
 \label{PBH-om}
\end{eqnarray}

where $T_0$ is the PBH temperature redshifted to the present time, $m_\text{BH}$ is the PBH mass, $N_\text{eff}$ is number of particle species with masses smaller than the BH temperature $T_\text{BH}$, and $z_\text{max}$ is the maximum redshift. The PBH temperature at present and the maximum redshift are given by \cite{Ejlli2011}

\begin{eqnarray}
\nonumber T_0 &=& 1.43\times 10^{13}\\
&\times& \sqrt{\left(\frac{100}{N_\text{eff}}\right)\left(\frac{100}{g_S\left(T\left(\tau_\text{BH}\right)\right)}\right)\left(\frac{m_\text{BH}}{1\,\text{g}}\right)}\quad \left(\text{Hz}\right)\nonumber\\
z_\text{max} &=& \left(\frac{32170}{N_\text{eff}}\right)^{2/3} \left(\frac{m_\text{BH}}{m_\text{Pl}}\right)^{4/3} \Omega_\text{P}^{1/3}  - 1,
\end{eqnarray}

where $\Omega_\text{P}$ is the density parameter of PBH at their production time and $m_\text{Pl}$ is the Planck mass and $g_S\left(T\left(\tau_{\rm BH}\right)\right)$ is the number of species contributing to the entropy of the primeval plasma at temperature $T\left(\tau_{\rm BH}\right)$ at the evaporating life-time $\tau_{\rm BH}$ \cite{Hawking:1976de,DonPage:1976} . The number of PBHs that take part in this process is included in the density parameter $\Omega_\text{P}$, see Ref. \cite{Ejlli2011} for details.

Now in order to extract the characteristic amplitude due to the stochastic background of GWs due to PBH evaporation we need an expression which connects $h_c$ to the density parameter $h_0^2\Omega_\text{gw}$. By using the definition of the density parameter in \eqref{Omega-i} and the expression for the energy density of GWs in \eqref{GW-en-dens}, we get
\begin{equation}\label{sto-dens-par}
h_c\left(0, f\right)\simeq 1.3\times 10^{-18} \sqrt{h_0^2\Omega_\text{gw}\left(f; t_0\right)} \left(\frac{1\,\text{Hz}}{f} \right).
\end{equation}
Now by using expression \eqref{PBH-om} into expression \eqref{sto-dens-par}, we get the following expression for the characteristic amplitude of GWs due to PBH evaporation

\begin{eqnarray}
\nonumber h_c\left(0, f\right) &=& 4.8 \times10^{-47} \left(\frac{N_\text{eff}}{100}\right) \left( f \over 1~{\rm Hz}\right) \left(1~ \text{g} \over m_{\rm BH}\right)\\
&\times&\sqrt{\int_0^{z_{\rm m}} \frac{\sqrt{1+z}}{e^{\left(\frac{2\pi f \left(1+z\right)}{T_0}\right)} -1} \, {\rm d}z}.
\label{eq:BH-evap}
\end{eqnarray}

In order to have an overview of the upper limit derived and the perspective to detect UHF GWs, in Fig.~\ref{fig:BH-Evap} is shown: the upper limits derived in the previous section, the estimated minimum detectable amplitude for the ALPS IIc and JURA considering the photo detector dark count rate, the maximum GWs amplitude generated in the production cavity, the estimated GWs amplitude for $N_{\rm eff}=100$, $\Omega_p=10^{-7}$ and BH masses ($10^{-3}$, $10^4$, $10^8$)~g, the prediction of the GWs from the sun and the nucleosynthesis upper limit, $\Omega_{\rm gw} \approx 10^{-5}$ \cite{Maggiore:1999vm}. The sensitivity to GW detection for ALPS IIc and JURA could reach better results for longer integration time, for example, $T=10^6-10^7$~s. A straightforward method, to integrate in time, is to modulate the field amplitude. In such a situation, the signal-to-noise ratio improves as $\sqrt{T}$. An alternative method, without the signal modulation, is to correlate the data stream from two different photodetectors. The electromagnetic wave is generated inside the FP cavity, and the transmission is on both mirrors of the cavity which are placed on the sides of the magnet. Having two photodetectors mounted on both sides of the magnet and correlating their signal in time would allow lowering the statistical noise of the detector. So, the time integration would let to a further gain in the sensitivity amplitude $h_c^{\rm min}$. Ultimately the ALPS IIc would be able to explore new amplitude regions of GWs which target source could be the predicted GWs generated from the evaporation of PBHs.

\section{Conclusions}
\label{sec:6}

A broad spectrum of the emission of GWs is predicted to exist in the universe, and some sources could generate GWs with frequencies higher than THz. The predicted conversion of gravitons into photons, due to the propagation in a static magnetic field, is not out of reach for current technologies. The technique of various detectors having the aim of counting single photons, at a narrow wavelength in a static magnetic field has been developed as detectors for the measurement of WISPs, a dark matter candidate, decaying to photons in the transverse magnetic field. Though the WISPs detection setups were not particularly designed to detect GW conversion, the generation of electromagnetic radiation as GWs propagate in a static magnetic field provides the possibility of using the published data, currently for the ALPS, OSQAR, CAST collaborations, to set the first upper limits on the amplitude of isotropic Ultra-High-Frequency GWs. We exclude the detection of GWs down to an amplitude $h_c^{\rm min}\approx6\times 10^{-26}$~at~$(2.7 - 14)\times10^{14}$~Hz and $h_c^{\rm min}\approx 5\times 10^{-28}$~at~$(5 - 12)\times10^{18}$~Hz at 95\% confidence level. Many theoretical potential ultra-high-frequency GW sources could be searched for using such similar experimental setups. The next generation experiments, such as the ALPS IIc and JURA facilities, or similar experiments using high static magnetic fields, are potential detectors for the graviton to photon conversion as well. The predicted ALPS IIc data taking or eventually JURA  will be able to produce more stringent upper limits on the amplitude of the stochastic wave background of GWs generated from PBH evaporation models. 

\section{Acknowledgement}
We are grateful to Prof. Bernard Schutz for helpful comments on the manuscript, and we recognise the support from the Leverhulme Emeritus Fellowship EM 2017-100.

\begin{appendix}

\section{Propagation of GWs in a constant magnetic field}\label{Appendix:A}

Here we review the graviton-photon mixing in a static external magnetic field, which can result in the conversion of gravitational waves into photons, the process which we describe in the main paper operating in the detectors designed to detect WISPs. In this section we closely follow Ref. \cite{Ejlli:2018hke}. To start with it is necessary first to write the total Lagrangian density $\mathcal L$ of the graviton-photon system. In our case, it is given by the sum of the following terms

 \begin{equation}\label{tot-lang}
 \mathcal L=\mathcal L_\text{gr}+\mathcal L_\text{em},
\end{equation}
where $\mathcal L_\text{gr}$ and $\mathcal L_\text{em}$ are respectively the Lagrangian densities of the gravitational and electromagnetic fields. These terms are respectively given by
\begin{eqnarray}\label{lag-dens}
L_\text{gr} &=&\frac{1}{\kappa^2}\,R,\\
\nonumber L_\text{em}&=&-\frac{1}{4}F_{\mu\nu}F^{\mu\nu}-\frac{1}{2}\int d^4x^\prime A_\mu\left(x\right) \Pi^{\mu\nu}\left(x, x^\prime\right)A_\nu\left(x^\prime\right),
\end{eqnarray}
where $R$ is the Ricci scalar, $g$ is the metric determinant, $F_{\mu\nu}$ is the total electromagnetic field tensor, $\kappa^2 \equiv16\pi G_\text{N}$ with $G_\text{N}$ being the Newtonian constant and $\Pi^{\mu\nu}$ is the photon  polarisation tensor in a magnetised medium.

The equations of motion from \eqref{tot-lang} and \eqref{lag-dens} for the propagating electromagnetic and GW fields components, $A^\mu$ and $h_{ij}$, in an external magnetic field are given by \cite{Ejlli:2018hke}

\begin{eqnarray}
\nonumber\nabla^2 A^0 &=& 0,\\ 
\nonumber \Box \boldsymbol A^i &+&\int d^4x^\prime \Pi^{i j}\left(x, x^\prime\right)\boldsymbol A_j\left(x^\prime\right)+ \partial^i\partial_\mu A^\mu= \\
&=&\kappa\,\partial_\mu[h^{\mu\beta}\bar {F}_{\beta}^{i}-h^{i\beta}\bar {F}_{\beta}^{\mu}],\nonumber\\
 \Box h_{ij} &=&-\kappa\, \left(B_i\bar B_j+\bar B_i B_j +\bar B_i\bar B_j\right),
 \label{system1}
\end{eqnarray}

where $A^\mu=\left(\phi, \boldsymbol A\right)$ is the incident electromagnetic vector-potential with magnetic field components $B_i$ and $\bar B_i$ are the components of the external magnetic field vector $\bar{\boldsymbol B}$. In obtaining the system \eqref{system1} we made use of the TT-gauge conditions for the GWs tensor $h_{0\mu}=0, h_i^i=0$ and $\partial^jh_{ij}=0$. As shown in details in Ref. \cite{Ejlli:2018hke}, the system \eqref{system1} can be linearized by making use of the slowly varying envelope approximation (SVEA) which is a WKB-like approximation for a slowly varying external magnetic field of spacetime coordinates. In this approximation, and for propagation along the observer's $\hat{\boldsymbol z}$ axis in a given cartesian coordinate system, equations \eqref{system1} can be written as \cite{Ejlli:2018hke}

\begin{equation}\label{schr-eq}
\left(\omega + i\partial_z\right)\Psi\left(z, \omega, \hat{\boldsymbol z}\right)\boldsymbol I+M\left(z, \omega\right) \Psi\left(z, \omega, \hat{\boldsymbol z}\right)=0, 
\end{equation}

where in \eqref{schr-eq} $\boldsymbol I$ is the unit matrix, {\small $\Psi\left(z, \omega, \hat{\boldsymbol z}\right)$ = $\left( h_\times, h_+,  A_x,  A_y\right)^\text{T}$ }is a four component field with $h_{\times, +}$ being the usual GW cross and plus polarisation states and $A_{x, y}$ are the usual propagating transverse photon states. In \eqref{schr-eq} $M\left(z, \omega\right)$ is the mass mixing matrix which is given by

\begin{equation}\label{mixing-matrix}
 M\left(z, \omega\right)=\begin{pmatrix}
  0 & 0 & -iM_{g\gamma}^x & iM_{g\gamma}^y \\
0 & 0 & iM_{g\gamma}^y & iM_{g\gamma}^x \\
iM_{g\gamma}^x & -iM_{g\gamma}^y & M_x & M_\text{CF} \\
-iM_{g\gamma}^y & -iM_{g\gamma}^x & M_\text{CF}^* & M_y
   \end{pmatrix},
   \end{equation}
   
where the elements of the mass mixing matrix $M$ are given by:
\begin{eqnarray}
\nonumber M_{g\gamma}^{x}&=&\kappa\,k \bar B_x/\left(\omega+k\right), \\
\nonumber M_{g\gamma}^y &=& \kappa \,k \bar B_{y}/\left(\omega+k\right), \\ 
\nonumber M_x&=&-\Pi_{xx}/\left(\omega+k\right), \\ 
\nonumber M_y&=&-\Pi_{yy}/\left(\omega+k\right), \\
\nonumber M_\text{CF} & = &- \Pi_{xy}/\left(\omega+ k\right).
\end{eqnarray}
Here $M_\text{CF}$ is a term which includes a combination of the Cotton-Mouton and Faraday effects in a plasma and which depends on the magnetic field direction with respect to the photon propagation.  Here $\omega$ is the total energy of the fields, namely $\omega=\omega_\text{gr}=\omega_{\gamma}$. In this work all the particles participating in the mixing are assumed to be relativistic, namely $\omega+k\simeq 2k$ with $k$ being the magnitude of the photon and graviton wave-vectors.

The system of differential equations \eqref{schr-eq} does not have closed solutions in the case when the mixing occurs in arbitrary matter that evolves in space and time, namely in the case when the system of differential equations is with variable coefficients such as in cosmological scenarios. However, in the case of mixing in a laboratory magnetic field, the system \eqref{schr-eq} can be simplified by considering a specific propagation of GWs  with respect to the magnetic field direction and by considering the propagation in the magnetic field without gas or a plasma present (see below). For example, first one can choose the magnetic field to be transverse to the photon direction of propagation such as $\bar {\boldsymbol B}=\left(\bar B_x, 0, 0\right)$ where we have $M_{g\gamma}^y=0$ and $M_\text{CF}=0$. Second, if there is a gas or a plasma in addition to the external magnetic field, usually we have that $M_x\neq M_y$ which essentially means that the transverse photon states have different indexes of refraction. In the case when one is able to achieve almost a pure vacuum in the laboratory, the contribution of the gas or plasma to the index of refraction can be safely neglected while there is also still present a contribution to the index of refraction due to the vacuum polarisation in the magnetic field. However, the latter contribution is completely negligible because the magnitude of the laboratory magnetic field is usually few Teslas and consequently is too small to have any appreciable effect on $M_x$ and $M_y$. 

As discussed above, let us consider first the case when the external magnetic field is completely transverse with respect to the photon direction of propagation where $M_{g\gamma}^y=0$ and $M_\text{CF}=0$. The fact that $M_\text{CF}=0$ is because $\bar B_y=0, \bar B_z=0$ and consequently the term corresponding to the Faraday effect is absent since this effect occurs only when the magnetic field has a longitudinal component along the electromagnetic wave direction of propagation. In addition, in $M_\text{CF}$ term it is also zero the term corresponding to the Cotton-Mouton effect in plasma because by convention we have chosen that $\bar B_y=0$. After these considerations several terms in the mixing matrix $M\left(z, \omega\right)$ are zero and in the case of the medium in the laboratory being homogeneous in space (including the magnetic field), then the mass mixing matrix $M$ does not depend on the coordinate $z$. In this case the commutator $[M\left(z, \omega\right), M\left(z^\prime, \omega\right)]=0$ and the solution of the system \eqref{schr-eq} is given by taking the exponential of $M\left(z, \omega\right)$. Consequently, we obtain the following solutions for the fields

\begin{footnotesize}
\begin{eqnarray}
\label{solutions-eq-motion}
h_\times\left(z, \omega, \hat{\boldsymbol z}\right)&=& \left[\cos\left(\Delta_x z\right) - i \frac{M_x \sin\left(\Delta_x z\right)} {2\Delta_x}\right] e^{i \left(\omega+\frac{M_x}{2}\right)z}\,h_\times\left(0, \omega, \hat{\boldsymbol z}\right) \nonumber\\
&+& \frac{M_{g\gamma}^x \sin\left(\Delta_x z\right)} {\Delta_x} e^{i \left(\omega+M_x/2\right)z} A_x\left(0, \omega, \hat{\boldsymbol z}\right),\nonumber\\
h_+\left(z, \omega, \hat{\boldsymbol z}\right)&=& \left[\cos\left(\Delta_y z\right) - i \frac{M_y \sin\left(\Delta_y z\right)} {2\Delta_y}\right] e^{i \left(\omega+\frac{M_y}{2}\right)z}\,h_+\left(0, \omega, \hat{\boldsymbol z}\right) \nonumber\\
&-& \frac{M_{g\gamma}^x \sin\left(\Delta_y z\right)} {\Delta_y} e^{i \left(\omega+M_y/2\right)z} A_y\left(0, \omega, \hat{\boldsymbol z}\right),\nonumber\\
A_x\left(z, \omega, \hat{\boldsymbol z}\right)&=& - \frac{M_{g\gamma}^x \sin\left(\Delta_x z\right)} {\Delta_x} e^{i \left(\omega+\frac{M_x}{2}\right)z}\,h_\times\left(0, \omega, \hat{\boldsymbol z}\right)\nonumber\\
& +& \left[\cos\left(\Delta_x z\right) + i \frac{M_x \sin\left(\Delta_x z\right)} {2\Delta_x}\right] e^{i \left(\omega+M_x/2\right)z}\, A_x\left(0, \omega, \hat{\boldsymbol z}\right),\nonumber\\
A_y\left(z, \omega, \hat{\boldsymbol z}\right)&=&  \frac{M_{g\gamma}^x \sin\left(\Delta_y z\right)} {\Delta_y} e^{i \left(\omega+\frac{M_y}{2}\right)z}\,h_+\left(0, \omega, \hat{\boldsymbol z}\right)\nonumber\\
& +& \left[\cos\left(\Delta_y z\right) + i \frac{M_y \sin\left(\Delta_y z\right)} {2\Delta_y}\right] e^{i \left(\omega+\frac{M_y}{2}\right)z}\, A_y\left(0, \omega, \hat{\boldsymbol z}\right),
\end{eqnarray}
\end{footnotesize}

where $h_{\times, +}\left(0, \omega, \hat{\boldsymbol z}\right)$ and $A_{x, y}\left(0, \omega, \hat{\boldsymbol z}\right)$ are respectively the GW and electromagnetic wave initial amplitudes at the origin of the coordinate system $z=0$, namely the amplitudes before entering the region where the magnetic field is located. In addition, we have defined
\begin{equation}\label{delta-def}
\Delta_{x, y} \equiv \frac{\sqrt{M_{x, y}^2 + 4 \left(M_{g\gamma}^{x}\right)^2}}{2}.
\end{equation}

\section{Electromagnetic energy fluxes generated with laboratory graviton-photon mixing }\label{Appendix:B}

In the previous appendix we found the solutions of the linearised equations of motion \eqref{schr-eq} for the GW fields $h_{\times, +}$ and for the electromagnetic wave fields $A_{x, y}$. In this section we use these solutions to find the energy flux of the electromagnetic radiation generated in the laboratory for the graviton-photon mixing (in equations abbreviated as graph). Before proceeding further is important to stress that in \eqref{solutions-eq-motion}, the GW amplitudes $h_{\times, +}$ are not dimensionless, as they are commonly defined in some textbooks, but have energy dimension units. This is due to the fact that in Ref. \cite{Ejlli:2018hke} the metric tensor is expanded as $g_{\mu\nu}=\eta_{\mu\nu} + \kappa h_{\mu\nu}$ where the GW tensor $h_{\mu\nu}$ has the physical dimensions of an energy. But in many cases one also writes $g_{\mu\nu}=\eta_{\mu\nu} + h_{\mu\nu}$ where in this case $h_{\mu\nu}$ is a dimensionless quantity. Since the latter case is quite common in the theory of GWs and because we want to conform to the literature, in expression \eqref{solutions-eq-motion} one has to simply replace $h_{\times, +}\left(0, t\right) = \tilde h_{\times, +}\left(0, t\right)/\kappa$, where $\tilde h_{\times, +}$ are now dimensionless amplitudes.

Consider the case when GWs enter a region of constant external magnetic field in the laboratory and that initially there are no electromagnetic waves. 

The assumption of no initial electromagnetic waves means that $A_x\left(0, \omega,  \hat{\boldsymbol z}\right)$ = $A_y\left(0, \omega, \hat{\boldsymbol z}\right)$=0 in the solutions \eqref{solutions-eq-motion}. Therefore, the expressions for the electromagnetic field components, in the graviton to photon mixing for a transverse propagation with respect to magnetic field, are given by

\begin{eqnarray}\label{photon-fields}
\nonumber A_x\left(z, \omega, \hat{\boldsymbol z}\right)&=& - \frac{M_{g\gamma}^x \sin\left(\Delta_x z\right)} {\kappa \Delta_x} e^{i \left(\omega+M_x/2\right)z}\,\tilde h_\times\left(0, \omega, \hat{\boldsymbol z}\right),\\
A_y\left(z, \omega, \hat{\boldsymbol z}\right)&=&  \frac{M_{g\gamma}^x \sin\left(\Delta_y z\right)} {\kappa\Delta_y} e^{i \left(\omega+M_y/2\right)z}\,\tilde h_+\left(0, \omega, \hat{\boldsymbol z}\right)
\end{eqnarray}

The expressions for the electromagnetic field components in \eqref{photon-fields}, even though very important, are not much useful for practical purposes since we usually detect electromagnetic radiations through their transported energy to the detector. For this reason is better to work with the Stokes parameter $I_\gamma\left(z, t\right)\equiv \Phi_\gamma\left(z, t\right)$ of the electromagnetic radiation generated in the graviton to photon mixing and which quantifies the energy flux (or energy density) of the electromagnetic radiation. The Stokes $\Phi_\gamma$ parameter, at a given point in space $z$, is defined as 

\begin{equation}\label{I-stokes}
\Phi_\gamma\left(z, t\right) \equiv \langle |E_x\left(z, t\right)|^2\rangle + \langle |E_y\left(z, t\right)|^2 \rangle,
\end{equation}
where $E_x$ and $E_y$ are the components of the electric field of electromagnetic radiation and the symbol $\langle \left(\cdot\right) \rangle$ denotes temporal average of quantities over many oscillation periods of electromagnetic radiation. The components of the electric field $\boldsymbol E$ are related to that of the vector-potential $\boldsymbol A$ through the relation $ E_{x, y}\left(z, t\right)=-\nabla A^0\left(z, t\right)-\partial_t A_{x, y}\left(z, t\right)$. For a globally neutral medium (if there is one except the magnetic field) in the laboratory we can choose $A^0=0$ from the first equation in \eqref{system1} and after we simply get $ E_{x, y}\left(z, t\right)=-\partial_t A_{x, y}\left(z, t\right)$.

In order to calculate the energy density of the electromagnetic radiation and related quantities in the graviton to photon mixing, we need first to make some assumptions on the GW signal which interacts with the magnetic field in the laboratory. In this work we concentrate on our study on a stochastic background of GWs with astrophysical or cosmological origin. It is rather natural to assume that the stochastic background of GWs is isotropic, unpolarized and stationary \cite{Allen:1997ad,Maggiore:1999vm}. In order to make more clear what these assumptions mean, we write the  GW amplitude tensor $\tilde h_{ij}\left(z, t\right)$ at $z=0$ as a Fourier integral

\begin{eqnarray}
\nonumber\tilde h_{ij}\left(0, t\right) &=&\sum_{\lambda=\times, +} \int_{-\infty}^{+\infty} \frac{d\omega}{2\pi}\int_{S^2} d^2\hat{\boldsymbol{n}}\,\tilde h_{\lambda}\left(0, \omega, \hat{\boldsymbol n}\right) e_{ij}^\lambda\left(\hat{\boldsymbol n}\right) e^{-i\omega t},\\& &\left(i, j=x, y, z\right),
\label{h-fou}
\end{eqnarray}
where $\hat{\boldsymbol n}$ is a unit vector on the two sphere $S^2$ which denotes an arbitrary direction of propagation of the GW, $d^2\hat{\boldsymbol n}=d\left(\cos\theta\right)d\phi$, $\lambda$ is the polarisation index of the GW with the usual cross and plus polarisation states and $e_{ij}^\lambda\left(\hat{\boldsymbol n}\right)$ is the GW polarisation tensor which has the property $e_{ij}^\lambda\left(\hat{\boldsymbol n}\right) e_{\lambda\prime}^{ij}\left(\hat{\boldsymbol n}\right)=2\delta_{\lambda\lambda^\prime}$. The assumptions that the stochastic background is isotropic, unpolarised and stationary means that the ensemble average of the Fourier amplitudes satisfies

\begin{eqnarray}\label{ens-av}
\nonumber \langle \tilde h_\lambda\left(0,\omega, \hat{\boldsymbol n}\right) \tilde h_{\lambda^\prime}^*\left(0, \omega^\prime, \hat{\boldsymbol n}^\prime\right)\rangle &=& 2\pi \delta\left(\omega-\omega^\prime\right) \frac{\delta^2\left(\hat{\boldsymbol n}, \hat{\boldsymbol n}^\prime\right)}{4\pi} \delta_{\lambda\lambda^\prime} \\
&\times& \frac{H\left(0, \omega\right)}{2},
\end{eqnarray}

where $H\left(0, \omega\right)$ is defined as the spectral density of the stochastic background of GW and it has the physical dimensions of Hz$^{-1}$ and $\delta^2\left(\hat{\boldsymbol n}, \hat{\boldsymbol n}^\prime\right)=\delta\left(\phi-\phi^\prime\right)\delta\left(\cos\theta-\cos\theta^\prime\right)$ is the covariant Delta function on the two sphere. One can check by using \eqref{h-fou} and \eqref{ens-av}, that the ensemble average $\langle \tilde h_{ij}\left(0, t\right) \tilde h^{ij}\left(0, t\right) \rangle$, is given by
 
\begin{eqnarray}\label{ens-av-1}
\langle \tilde h_{ij}\left(0, t\right) \tilde h^{ij}\left(0, t\right)\rangle  &=& 2\int_{-\infty}^{+\infty} \frac{d\omega}{2\pi} H\left(0, \omega\right)\\
\nonumber&=&4\int_{0}^{+\infty} \frac{d\left(\log\omega\right)}{2\pi}\, \omega\, H\left(0, \omega\right)\nonumber \\
&\equiv& 2\int_0^{+\infty} d\left(\log\omega\right) h_c^2\left(0, \omega\right),
\end{eqnarray}
where the last equality in  \eqref{ens-av-1} defines the characteristic amplitude, $h_c$ (dimensionless), of a stochastic background of GWs. In obtaining \eqref{ens-av-1} we used the fact that for an unpolarised stochastic background, we have that on average, $\langle |\tilde h_\times\left(0, \omega\right)|^2  \rangle=\langle |\tilde h_+\left(0, \omega\right)|^2 \rangle \neq 0$ while the ensemble average of the mixed terms vanish identically. We may observe that by comparing the two last equalities in \eqref{ens-av-1} we get $h_c^2\left(0, \omega\right)= 2 \omega H\left(0, \omega\right)/\left(2\pi\right)$.

With the expressions \eqref{h-fou}-\eqref{ens-av-1} in hand we are at the position to calculate $\Phi_\gamma$ and relate it with $h_c$ or $H$. Let us at this point write the components of the vector-potential for $\hat{\boldsymbol{n}}=\hat{\boldsymbol{z}}$ as Fourier integrals

\begin{equation}\label{vect-pot-exp}
A_{x, y}\left( z, t\right)=\int_{-\infty}^{+\infty} \frac{d\omega}{2\pi}\,\int d^2\hat{\boldsymbol{z}}\, A_{x, y}\left(z, \omega, \hat{\boldsymbol{z}}\right) e^{-i\omega t}.
\end{equation}
The by using expression \eqref{vect-pot-exp} in $E_{x, y}\left(z, t\right)=-\partial_t A_{x, y}\left(z, t\right)$ and then putting all together in the expression of the Stokes parameter \eqref{I-stokes}, we get

\begin{eqnarray}
\Phi_\gamma\left(z, t\right)&=&\left(M_{g\gamma}^x\right)^2 \int_0^{+\infty} \frac{d\omega}{2\pi} \times\\
\nonumber &\times& \left[ \frac{ \sin^2\left(\Delta_x z\right)} {\Delta_x^2} + \frac{  \sin^2\left(\Delta_y z\right)} {\Delta_y^2} \right] \frac{\omega^2 H\left(0, \omega\right)}{\kappa^2},
\label{I-stokes-1}
\end{eqnarray}

where in obtaining the expression \eqref{I-stokes-1} we used also expression \eqref{ens-av}. In addition, we may note that both $\Delta_x$ and $\Delta_y$ implicitly depend on $\omega$ through $M_x$ and $M_y$ and thus explain the reason why $\Delta_{x, y}$ do appear under the integral sign in \eqref{I-stokes-1}. On the other hand, $M_{g\gamma}^x$ does not depend on $\omega$ since we are considering relativistic particles with $\omega\simeq k$. 

Now in order to relate the total energy density of the formed electromagnetic radiation in the graviton-photon mixing, we may note from expression \eqref{I-stokes-1} that the energy density contained in a logarithmic energy interval, is given by

\begin{footnotesize}
\begin{eqnarray}\label{en-log-int}
\frac{d\Phi_\gamma^\text{graph}\left(z, \omega; t\right)}{d\left(\log\omega\right)} &=& \left(M_{g\gamma}^x\right)^2 \left[ \frac{ \sin^2\left(\Delta_x z\right)} {\Delta_x^2} + \frac{  \sin^2\left(\Delta_y z\right)} {\Delta_y^2} \right] \frac{\omega^3 H\left(0, \omega\right)}{2\pi \kappa^2}\nonumber\\
\nonumber&=& \frac{\left(M_{g\gamma}^x\right)^2}{2} \left[ \frac{ \sin^2\left(\Delta_x z\right)} {\Delta_x^2} + \frac{  \sin^2\left(\Delta_y z\right)} {\Delta_y^2} \right] \frac{\omega^2 h_c^2\left(0, \omega\right)}{\kappa^2}.\\
&&
\end{eqnarray}
\end{footnotesize}
The expression $d\Phi_\gamma^\text{graph}\left(z, \omega; t\right)/d\left(\log\omega\right)$ in \eqref{en-log-int} is an expression which tells us how much of the total energy density is contained in a logarithmic energy interval. The expression \eqref{en-log-int} can be written also in an equivalent form in terms of the density parameter of photons $\Omega_\gamma$ and the density parameter of GWs, $\Omega_\text{gw}$. The density parameter of a specie $i$ of particles at the energy $\omega$ is defined as

\begin{equation}\label{Omega-i}
\Omega_i\left(z, \omega; t\right) \equiv \frac{1}{\rho_c}\frac{d\rho_i\left(z, \omega; t\right)}{d\left(\log \omega\right)},
\end{equation}
where $\rho_c$ is the critical energy density to close the universe, $\rho_c=6 H_0^2/\kappa^2$, where $H_0$ is the Hubble parameter $H_0=100\, h_0$ (km/s/Mpc) with $h_0$ being a dimensionless parameter which is determined experimentally. In addition since the energy density (or energy flux $\Phi$) of GWs is given by
 
\begin{eqnarray}\label{GW-en-dens}
\nonumber \rho_\text{gw}\left(0, t\right)&=&\frac{\langle \dot h_{ij}\left(0, t\right) \dot h^{ij}\left(0, t\right)\rangle}{2\kappa^2}\\&=& \int_0^{+\infty} d\left(\log\omega\right) \frac{\omega^2 h_c^2\left(\omega\right)}{\kappa^2},
\end{eqnarray}

where we used \eqref{h-fou}-\eqref{ens-av}, we have from \eqref{en-log-int}, \eqref{Omega-i} and \eqref{GW-en-dens} that 

\begin{eqnarray}\label{dens-par-osc}
\nonumber h_0^2\Omega_\gamma^\text{graph}\left(z, \omega; t\right) &=& \frac{\left(M_{g\gamma}^x\right)^2}{2} \left[ \frac{ \sin^2\left(\Delta_x z\right)} {\Delta_x^2} + \frac{  \sin^2\left(\Delta_y z\right)} {\Delta_y^2} \right]\times\\
&\times& h_0^2 \Omega_\text{gw}\left(0, \omega; t\right).
\end{eqnarray}

The expressions \eqref{en-log-int} and \eqref{dens-par-osc} essentially give a complete description of how the graviton to photon mixing propagates in space in a transverse and constant magnetic field and uniform medium. Both \eqref{en-log-int} and \eqref{dens-par-osc} are equally important and can be used in different contexts in order to compare with experimental data. It is very important to analyze these expressions in some limiting cases. Consider the case when in the laboratory there is a medium (gas or plasma) in addition to the magnetic field and when $M_x=M_y$. The last condition essentially means that both propagating transverse states of the electromagnetic radiation have the same index of refraction. When $M_x=M_y$, we have that $\Delta_x=\Delta_y$ and consequently we get for \eqref{dens-par-osc} that 

\begin{equation}\label{dens-par-osc-1}
h_0^2\Omega_\gamma^\text{graph}\left(z, \omega\right) = \left(M_{g\gamma}^x\right)^2\left[ \frac{ \sin^2\left(\Delta_x z\right)} {\Delta_x^2} \right] h_0^2 \Omega_\text{gw}\left(0, \omega\right).
\end{equation}

Another important situation is  when in the laboratory there is not a medium but only a magnetic field in vacuum. In this case we have that $\Delta_x=\Delta_y=M_{g\gamma}^x$ and the graviton-photon mixing is maximal or resonant 

\begin{equation}\label{dens-par-osc-2}
h_0^2\Omega_\gamma^\text{graph}\left(z, \omega\right) = [\sin^2\left(M_{g\gamma}^x z\right)]  h_0^2 \Omega_\text{gw}\left(0, \omega\right).
\end{equation}
Expression \eqref{dens-par-osc-2} tell us that the graviton-photon mixing has an oscillatory behaviour with the distance $z$ in the maximum mixing case. For a long distance of travelling, there are values of $z$ which make $\sin^2\left(M_{g\gamma}^x z\right)=1$ and in that case we have that all GWs are transformed into electromagnetic waves. However, since $M_{g\gamma}^x$ is usually a very small quantity, one needs huge distances of travelling in order to achieve this situation. In many practical cases one has that $M_{g\gamma}^x z\ll 1$ and we can approximate $\sin^2\left(M_{g\gamma}^x z\right) \simeq \left(M_{g\gamma}^x z\right)^2$ in the maximum mixing case.\\

\end{appendix}

\end{document}